\documentclass[12pt]{article}
\usepackage{latexsym}
\usepackage{mathrsfs}
\usepackage{amssymb}
\usepackage{amsmath}
\usepackage{amsfonts, epsf,epsfig,color}
\usepackage{graphicx}
\usepackage{tensor}
\usepackage[all]{xy}
\include{macro}

\newcommand{\C}{\mathbb{C}}

\newcommand{\CP}{\mathbb{CP}}

\renewcommand{\P}{\mathbb{P}}

\newcommand{\M}{\mathbb{M}}

\newcommand{\T}{\mathbb{T}}

\newcommand{\p}{\partial}

\newcommand{\D}{\mathrm{D}}

\newcommand{\G}{\mathcal{G}}
\newcommand{\cO}{\mathcal{O}}

\renewcommand{\P}{\mathbb{P}}

\newcommand{\rd}{\, \mathrm{d}}

\newcommand{\be}{\begin{equation}\label}
\newcommand{\ee}{\end{equation}}
\newcommand{\bea}{\begin{eqnarray}\label}
\newcommand{\eea}{\end{eqnarray}}

\newcommand{\dbar}{\bar\p}

\begin{document}

\title{\textbf{The supersymmetric Penrose\\ transform in six dimensions}}

\author{L J Mason and R A Reid-Edwards\\
\\
The Mathematical Institute, University of Oxford\\
24-29 St.~Giles, Oxford OX1 3LB, United Kingdom\\
}

\date{}

\maketitle

\vspace{5cm}

\begin{abstract}
We give a supersymmetric extension to the  six-dimensional Penrose transform and give an integral formula for the on-shell $(0,2)$ supermultiplet. The relationship between super fields on space-time and twistor space is clarified and the space-time superfield constraint equations are derived from the geometry of supertwistor space. We also explain the extension to more general $(0,n)$ supermultiplets and give twistor actions for these theories.
\end{abstract}

\newpage

\section{Introduction}

Twistor methods have become powerful tools in the study of
supersymmetric gauge theory in four dimensions
\cite{Witten:2003nn,ArkaniHamed:2009dn,Adamo:2011pv}.  Much progress in
the study of 4-dimensional gauge theories also arises from higher
dimensional considerations, both via considerations of the $(0,2)$
theory in 6 dimensions and the AdS$_5$/CFT correpsondence.
It is natural  to ask whether these twistor techniques can be extended
to higher dimensional theories with large amounts of supersymmetry as
a first step to making contact with these higher dimensional
approaches.   This question is not just one of finding more efficient
methods of calculation; the character of twistor space is very
sensitive to space-time dimension and so it is possible that some of
the features of poorly understood theories, that only occur in higher
dimensions, may be more readily understood from a twistor
description. 

In \cite{Mason:2011nw,Saemann:2011nb} the
description of linearised six-dimensional conformal field theories in
twistor space was investigated.  This article is a
sequel to \cite{Mason:2011nw} and considers the extension of the results
of \cite{Mason:2011nw} to $(2,0)$ supersymmetry.  The relevant $(2,0)$ supertwistor
space was introduced in \cite{Harnad:1995zy,Chern:2009nt}, following on from the
work of \cite{Ferber:1977qx}; however, the twistor superfields and
Penrose transform presented in \cite{Chern:2009nt} does not correctly reproduce the
chiral superspace superfield \cite{Howe:1983fr} which is known to
correctly describe the massless linear $(2,0)$ supermultiplet. In this
article we present an
integral transform which does correctly reproduce the space-time
superfield. One of the notable features of the supertwistor
description is that remarkable simplicity of the twistor superfield;
the space-time superfield involves sixteen terms, many of which are
quite complicated, by contrast the twistor representative has only
five simple terms. Indeed, the $(2,0)$ twistor superfield is
analogous to the corresponding twistor representative which
describes the ${\cal N}=4$ super Yang-Mills multiplet in four
dimensions. Given the simplicity with which supeconformal theories
incorporating self-dual fields have been described by using twistor
geomery, it is hoped that twistor space might provide a more natural
framework in which to investigate more exotic theories such as the
six-dimensional $(2,0)$ superconformal theory.  We remark that the
$(1,1)$ six-dimensional super Yang-Mills theory has been discussed
from a six-dimensional
analogue of the four-dimensional ambitwistor perspective in \cite{Saemann:2012rr}. 

The outline of this paper is as follows. In the following section, we review some of the salient features of bosonic conformal field theories in terms of twistor variables, including descriptions of self-dual gerbes, chiral fermions and scalar fields. In section three we recall the definition of $(2,0)$ supersymmetry in terms of superfields on chiral superspace $\M$ and the supertwistor space introduced in \cite{Harnad:1995zy,Chern:2009nt}; in section four we define the direct Penrose transform from twistor superfields on $(2,0)$ supertwistor space ${\cal Q}$ to superfields on complexified chiral superspace $\M$. In section five we show that the content of the linearised $(2,0)$ theory can be encoded as an element of $H^2$ on supertwistor space. In section six we use the results of the previous section to construct first order twistor actions on supertwistor space. We outline how these constructions extend to massless linearised field theories with $(2,0)$ supersymmetry and also discuss the conjectured $(4,0)$ superconformal theory \cite{Hull:2000ih,Hull:2000rr} which is conjectured to describe an as yet poorly understood, non-geometric, model of gravity.

\section{Twistor geometry}

In this section we summarise features of the twistor description of
linearised bosonic conformal field theories in six-dimensional twistor
space\footnote{See \cite{Huggett:1986fs,Ward:1990vs} for an
  introduction to four-dimensional twistor geometry and
  \cite{Adamo:2011pv} for a review of recent applications to ${\cal
    N}=4$ Yang-Mills theory. Discussions of six-dimensional twistor
  geometry may also be found in \cite{Hughston:1986hb}.}. Points  in complexified space-time, $\C^6$, can be parametrized by $x^{AB}=-x^{BA}$ where $A,B=1,2,3,4$ are spinor indices.   In these coordinates the metric becomes
$$
\rd s^2=\frac 12 \varepsilon_{ABCD}\rd x^{AB} \rd x^{CD} \;,
$$
where $\varepsilon_{ABCD}=\varepsilon_{[ABCD]}$.  
 Twistor space is a 6-quadric $Q\subset\C\P^7$  given, in homogenous coordinates on $\C\P^7$
$$
Z=\left(\begin{array}{c}\omega^A\\
\pi_A
\end{array}\right)\qquad  A,B=1,2,3,4,
$$
by
$$
Z^2=\omega^A\pi_A=0\, .
$$
A point $x$ in space-time corresponds to a $\C\P^3\subset Q$, which we will  denote by $S_x$, determined by the incidence relation
\begin{equation}\label{Incidence}
\omega^A=x^{AB}\pi_B\, .
\end{equation}
Here we take $\pi_A$ to be homogeneous coordinates on the $\CP^3$, $S_x$.  Given some region $R$ in space-time (which could, for example be a real slice) we denote the region in twistor space, swept out by the $S_x$ for $x$ in the region $R$, as $Q'$.  

We will be interested in a certain family of conformally invariant massless fields $\phi_{A_1\ldots A_n}$ of spin $n/2$
$$
\Gamma_n=\{ \phi_{A_1\ldots A_n}=\phi_{(A_1\ldots A_n)} \mbox{ on } R\, |\nabla^{BA_1}\phi_{A_1\ldots A_n}=0\}\, .
$$
There are two ways of expressing such space-time fields on twistor
space: in terms of cohomology classes either in $H^3(Q',\cO(-n-4))$
or in $H^2(Q',\cO(n-2)$, where $n/2$ is the spin of the field in question
(we will abbreviate these to $H^3(-n-4)$ and $H^2(n-2)$).  Cohomology
classes in $H^p(r)$  can
be represented in a variety of ways: traditionally they are
represented by holomorphic functions
of homogeneity degree $r$ defined on intersections of $r+1$ sets of
some cover of a region of $Q$ by Stein neighbourhoods. However, here
we will consider them as $\dbar$-closed $(0,p)$-forms
modulo exact forms.

The relationship between these two descriptions of the same $\Gamma_n$
turns out to be
intimately tied up with the notion of extending these cohomology
classes off the quadric $Q$ into $\CP^7$ (i.e., to formal
neighbourhoods of $Q$ in $\CP^7$) \cite{Mason:2011nw}. 
Elements of $H^3(-n-4)$ may be extended off the quadric without
obstruction although there is much ambiguity in the extension at each
order. Nontrivial elements of $H^2(n-2)$ can be extended uniquely to
the $n-1^{st}$ formal neighbourhood but beyond this there is an
obstruction to extending it to the $n$th order precisely given by the
corresponding $H^3(-n-4)$. This relationship between the elements of
$H^3(-n-4)$ and elements of $H^2(n-2)$ can be expressed by
\cite{Mason:2011nw} 
\be{TT}
\bar{\p}c\sim (\omega\cdot\pi)^{n+1}g
\ee
where $c\in H^2(n-2)$ and $g\in H^3(-n-4)$.

The transform from elements of $H^3(-n-4)$  to space-time fields is
given explicitly by the direct Penrose transform given by the integral representation
$$
\phi_{A_1\ldots A_n}=\int_{S_x}\D^3\pi\;\pi_A\pi_B\,h(\omega,\pi)\, ,
\qquad
\D^3\pi:=\varepsilon^{ABCD}\pi_A\rd\pi_B\wedge\rd\pi_C\wedge\rd\pi_D\, ,
$$
where the integrand is restricted to the $S_x\simeq \CP^3\subset Q$ defined by the
incidence relation (\ref{Incidence}).
For the $(2,0)$ multiplet we will be most interested in the special cases
$$
H_{AB}(x)=\int_{S_x}\D^3\pi\;\pi_A\pi_B\,h(\omega,\pi)\;,
	\qquad	\Psi_A(x)=\int_{S_x}\D^3\pi\;\pi_A\,\psi(\omega,\pi)\;,	\qquad	\Phi(x)=\int_{S_x}\D^3\pi\,\phi(\omega,\pi)\;.
$$
 The space-time fields, defined in this way, automatically satisfy the
 equations of motion $\nabla^{AC}H_{CB}=0$, $\nabla^{AB}\Psi_B=0$ and
 $\Box\Phi=0$.  Elements of $H^2(n-2)$ may be related to their
 space-time counterparts by the potential, modulo gauge, argument
 given in the Appendix of \cite{Mason:2011nw} and also \cite{Saemann:2011nb} or by the
 relation \eqref{TT}.

\section{$(2,0)$ Supertwistor Geometry}

In this article we are primarily interested in extending the analysis
in \cite{Mason:2011nw} of linear six-dimensional conformally invariant
field theories to linear superconformal field theories. The supertwistor geometry presented in
\cite{Chern:2009nt,Ferber:1977qx} is reviewed in this section. To motivate the superspace-time, we review the space-time superfield description of the $(2,0)$ chiral supermultiplet  \cite{Howe:1983fr}.

\subsection{(2,0) Superspace}

The complexified $(2,0)$-superspace is $\C^{6|16}$ with coordinates $(x^{AB},\theta^{A\alpha})$ where $\alpha,\beta,...=1,2,3,4$ are indices for the R-symmetry group\footnote{This is the isometry group of the $S^4$ factor in the dual $AdS_7\times S^4$ theory.} Sp(2)$\simeq$SO(5) so that there is an invariant skew tensor $\Omega^{\alpha\beta}$. We can use $\Omega^{\alpha\beta}$ and its inverse $\Omega_{\alpha\beta}$, which satisfy $\Omega^{\alpha\beta}\Omega_{\beta\gamma}=\delta^\alpha_\gamma$ to raise and lower the R-symmetry indices.  Because $\Omega^{\alpha\beta}$  is skew, we must be careful about signs when we raise and lower indices and stick to the convention that $A^\alpha=\Omega^{\alpha\beta}A_\beta, A_\alpha=\Omega_{\alpha\beta}A^\beta$ noting that with this convention $\Omega_{\alpha\beta}\Omega^{\alpha\beta}=-4$.  

We recall the left and right acting covariant superspace derivatives
\begin{equation}\label{DQ}
{\cal D}^{\alpha}_A=-\Omega^{\alpha\beta}\frac{\p}{\p\theta^{A\beta}}+\theta^{B\alpha}\nabla_{AB}\,,	\qquad	Q^{\alpha}_A=\Omega^{\alpha\beta}\frac{\p}{\p\theta^{A\beta}}+\theta^{B\alpha}\nabla_{AB}\,,
\end{equation}
where
$$
\nabla_{AB}:=\frac{\p}{\p x^{AB}}.
$$
These satisfy the (anti)commutation relations
$$
\{Q^{\alpha}_A,Q^{\beta}_B\}=-2\Omega^{\alpha\beta}\nabla_{AB},	\qquad	\{{\cal D}^{\alpha}_A,Q^{\beta}_B\}=0	\qquad	\{{\cal D}^{\alpha}_A,{\cal D}^{\beta}_B\},=2\Omega^{\alpha\beta}\nabla_{AB}\, .
$$
The $Q^{\alpha}_A$ generate the (2,0) supersymmetry algebra and 
${\cal D}^\alpha_A$ the supersymmetrically-invariant operators.. 

The field content of the (2,0) theory consists of a self-dual gerbe curvature $H_{AB}(x)$ (with three on-shell degrees of freedom), eight fermions $\Psi_A^{\alpha}(x)$ (or `gerbinos') and five scalars  $\Phi^{\alpha\beta}(x)=-\Phi^{\beta\alpha}$, which satisfy $\Omega_{\alpha\beta}\Phi^{\alpha\beta}(x)=0$.  These can be encoded in a traceless scalar superfield   \cite{Howe:1983fr}
\begin{equation}\label{trace}
{\cal W}^{\alpha\beta}=-{\cal W}^{\beta\alpha}, \qquad
\Omega_{\alpha\beta}{\cal W}^{\alpha\beta}=0
\end{equation}
 satisfying the differential equation
\begin{equation}\label{constraint}
({\cal D}^{\alpha}_A{\cal W}^{\beta\gamma})_0
:={\cal D}^{\alpha}_A{\cal W}^{\beta\gamma}
+\frac{1}{5}\Omega_{\rho\sigma}{\cal D}^{\rho}_A\left(2\Omega^{\alpha\beta}{\cal W}^{\sigma\gamma}-2\Omega^{\alpha\gamma}{\cal W}^{\sigma\beta}+\Omega^{\beta\gamma}{\cal W}^{\sigma\alpha}\right)=0\,.
\end{equation}
This constrains the trace-free part $({\cal D}^{\alpha}_A{\cal
  W}^{\beta\gamma})_0$ of $D_A^{\alpha}{\cal
  W}^{\beta\gamma}$ to vanish, but the traces are unconstrained.  From the perspective of the conjectured $AdS_7/CFT$ correspondence, this constraint arises as a consistency condition on the embedding of an M5-brane in eleven-dimensional superspace.

Since ${\cal D}^{\alpha}_A$ and $Q^{\alpha}_A$ anticommute, ${\cal D}^{\alpha}_A$ maps a superfield to a superfield. We can therefore introduce the spin-half and spin-one superfields $\Xi_A^{\alpha}$ and ${\cal H}_{AB}$ respectively \cite{Howe:1983fr}. These are given in terms of the scalar superfield ${\cal W}^{\alpha\beta}(x,\theta)$ by
$$
\Xi_A^{\alpha}(x,\theta)=\frac{2}{5}\Omega_{\beta\gamma}{\cal D}_A^{\beta}{\cal W}^{\gamma\alpha}\,,
\qquad
{\cal H}_{AB}(x,\theta)=\frac{1}{4}\Omega_{\alpha\beta}{\cal D}_A^{\alpha}\Xi^{\beta}_B\,.
$$
The leading terms in the $\theta$-expansion of $\Xi_A^{\alpha}$ and ${\cal H}_{AB}$ are the spin-half $\Psi^{\alpha}_A$ and self-dual gerbe $H_{AB}$ fields respectively. These superfields may be simply combined into a superspace three-form
$$
{\cal H}=\frac{1}{6}{\cal H}_{AB}\,E^{AC}\wedge E_{CD}\wedge E^{BD}+\frac{1}{2}\Xi^{\alpha}_A\,E^{AB}\wedge E_{BC}\wedge E^C_{\alpha}+\frac{1}{2}{\cal W}^{\alpha\beta}\,E_{AB}\wedge E^A_{\alpha}\wedge E^B_{\beta}\;,
$$
where we have introduced the one-forms
$$
E^{AB}:=\rd x^{AB}+\Omega_{\alpha\beta}\theta^{B\alpha}\rd \theta^{A\beta}\;,	\qquad	E^A_{\alpha}:=\Omega_{\alpha\beta}\rd \theta^{A\beta}\;,
$$
so that
\begin{eqnarray}
\mathscr{D}:&=&\rd x^{AB}\nabla_{AB}+\rd \theta^{A\alpha}\frac{\p}{\p\theta^{A\alpha}}\nonumber\\
&=&E^{AB}\nabla_{AB}+E^A_{\alpha}{\cal D}_A^{\alpha}\nonumber\;.
\end{eqnarray}
Using these definitions and the constraint equation (\ref{constraint}), it is  simple task to show that the leading components $\Phi^{\alpha\beta}$, $\Psi^{\alpha}_A$, and $H_{AB}$ of the superfields ${\cal W}^{\alpha\beta}$, $\Xi^{\alpha}_A$ and ${\cal H}_{AB}$ respectively exhaust the possible independent components of the on-shell supermultiplet. One way of seeing this is to observe the closure of the relations \cite{Howe:1983fr}
$$
{\cal D}^{\alpha}_A{\cal W}^{\beta\gamma}=-\Omega^{\alpha\beta}\Xi^{\gamma}_A+\Omega^{\alpha\gamma}\Xi^{\beta}_A-\frac{1}{2}\Omega^{\beta\gamma}\Xi^{\alpha}_A\,,
\qquad
{\cal D}^{\alpha}_A\Xi^{\beta}_B=2\nabla_{AB}{\cal W}^{\alpha\beta}-\Omega^{\alpha\beta}{\cal H}_{AB}\,	,
$$
$$	
{\cal D}^{\alpha}_A{\cal H}_{BC}=\nabla_{AB}\Xi^{\alpha}_C+\nabla_{AC}\Xi^{\alpha}_B\,.
$$
The super-three form then satisfies
$$
\mathscr{D}{\cal H}=0\,.
$$
Thus it is natural to consider a super-two-form potential ${\cal B}$
such that ${\cal H}=\mathscr{D}{\cal B}$. In section five we shall see
how the superfield components of such a two-form arise naturally from twistor theory.

\subsection{(2,0) Supertwistors}

Reflecting the fact that six-dimensional twistors are the fundamental representation of the six-dimensional conformal group SO(2,6), we would like to define six-dimensional supertwistors to be in the fundamental representation of the six-dimensional superconformal group SO(2,6$|$4). A supertwistor space with this requirement was described by  \cite{Chern:2009nt,Ferber:1977qx}, in which the supertwistor space ${\cal Q}\subset \C\P^{7|4}$ has coordinates 
$$
{\cal Z}^I=\left(\begin{array}{c}\omega^A\\
\pi_A\\
\eta^{\alpha}
\end{array}\right)\;,
$$
where the $\eta^{\alpha}$ are four anti-commuting grassman coordinates
and $\alpha=1,2,3,4$ is an $R$-symmetry index. The superquadric condition
\be{SQ}
{\cal Z}^2=\omega^A\pi_A-\frac{1}{2}\Omega_{\alpha\beta}\eta^{\alpha}\eta^{\beta}=0\;,
\ee
then follows naturally from the incidence relations
\begin{equation}\label{Sincidence}
\omega^A=x^{AB}\pi_B+\frac{1}{2}\Omega_{\alpha\beta}\theta^{A\alpha}\theta^{B\beta}\pi_B\,,	\qquad	\eta^{\alpha}=\theta^{\alpha A}\pi_A\;.
\end{equation}
These incidence relations describe, for a fixed point in superspace $(x^{AB},\theta^{A\alpha})$, a $\C\P^3\subset {\cal Q}$ as in the bosonic case. Using the incidence relations (\ref{Sincidence}), the (2,0) generators can be described on twistor space by the linear operators (\ref{DQ})
\begin{equation}\label{Q-twis}
 Q^{\alpha}{}_A=-\Omega^{\alpha\beta}\pi_A\frac{\p}{\p\eta^{\beta}}+\eta^{\alpha}\frac{\p}{\p \omega^A}\,.
\end{equation}
A twistor description of the remaining generators of the superconformal algebra were found, for example, in \cite{Chern:2009nt}. These generators preserve the superquadric \eqref{SQ} as expected.  We will see later that the operators ${\cal D}^{\alpha}_A$, which play a key role in the direct Penrose transform, do not directly descend to operators on $(2,0)$ supertwistor space. 

\subsection{Twistor Superfields}

Now that we have the basic geometric set-up, we can consider supersymmetric field theories on this space constructed from elements of $H^3(-n-4)$ and $H^2(n-2)$. In this section we will consider twistor superfields constructed from $H^3(-n-4)$ representatives, leaving the discussion of $H^2$ twistor superfields to section five. The most general $H^3$ twistor superfield can be expanded in powers of the anti-commuting $\eta^{\alpha}$ to give
\begin{eqnarray}\label{sf}
G(\omega,\pi,\eta)&=&\tilde{h}(\omega,\pi)+\tilde{\psi}_{\alpha}(\omega,\pi)\,\eta^{\alpha}+\frac{1}{2}S_{\alpha\beta}(\omega,\pi)\,\eta^{\alpha}\,\eta^{\beta}+\frac{1}{3!}\,\psi^{\alpha}(\omega,\pi)\,\varepsilon_{\alpha\beta\lambda\rho}\,\eta^{\beta}\,\eta^{\lambda}\,\eta^{\rho}\nonumber\\
&&+\frac{1}{4!}\,h(\omega,\pi)\,\varepsilon_{\alpha\beta\lambda\rho}\,\eta^{\alpha}\,\eta^{\beta}\,\eta^{\lambda}\,\eta^{\rho}\nonumber\;,
\end{eqnarray}
where we require that the bosonic coefficients all take values in $H^3(Q;{\cal O}(-n-4))$, each according to the spin $n/2$ of the space-time fields they describe. The superfield then has overall projective weight $-2$. 

An important feature is that the cohomology groups $H^3(Q',\cO(-2))=H^3(Q',\cO(-3))=0$ \cite{Baston:1989vh} so that the first two terms will not contribute to the on-shell degrees of freedom of the superfield.  They nevertheless have to be present in any off-shell description if we wish to make manifest the full supersymmetry.

 The fact that $\tilde{h}$ and $\tilde{\psi}_{\alpha}$ carry no on-shell physical information, and may therefore be omitted, simplifies the form of the superfield. Further simplification may be seen by considering the R-symmetry decomposition: $\underline{6}\rightarrow \underline{5}\oplus\underline{1}$, of the weight $-4$ field $S^{\alpha\beta}$ in $G$
$$
S^{\alpha\beta}(\omega,\pi):=\phi^{\alpha\beta}(\omega,\pi)+\frac{1}{4}\Omega^{\alpha\beta} \,{\cal S}(\omega,\pi)\;,
$$
where $\Omega_{\alpha\beta}\phi^{\alpha\beta}(\omega,\pi)=0$. The superfield $G(\omega,\pi,\eta)$ then includes the terms $ \frac{1}{2}\,\phi_{\alpha\beta}(\omega,\pi)\,\eta^{\alpha}\,\eta^{\beta}+\frac{1}{8}\,{\cal S}(\omega,\pi)\,\eta^{\alpha}\eta_{\alpha}$. Restricting the physical on-shell fields to lie on the superquadric ${\cal Z}^2=0$, the on-shell superfield includes the terms $\frac{1}{2}\,\phi_{\alpha\beta}(\omega,\pi)\,\eta^{\alpha}\,\eta^{\beta}+\breve{\cal S}(\omega,\pi)$, for a bosonic field $\breve{\cal S}(\omega,\pi)=-\frac{1}{4}\,{\cal S}(\omega,\pi)\,\omega^A\pi_A\in H^3(\P\T;{\cal O}(-2))$. Again, this cohomology is trivial and we can ignore the contribution given by $\breve{\cal S}$. The only physical scalar degrees of freedom are the \emph{five} representatives $\phi^{\alpha\beta}(\omega,\pi)$ which satisfy $\Omega_{\alpha\beta}\phi^{\alpha\beta}(\omega,\pi)=0$. The superfield (\ref{sf}) thus contains a lot of superfluous information and the physical content may be described equivalently by the truncated superfield
$$
G(\omega,\pi,\eta)=\frac{1}{2}\phi_{\alpha\beta}(\omega,\pi)\,\eta^{\alpha}\,\eta^{\beta}+\frac{1}{3!}\,\psi^{\alpha}(\omega,\pi)\,\varepsilon_{\alpha\beta\lambda\rho}\,\eta^{\beta}\,\eta^{\lambda}\,\eta^{\rho}+\frac{1}{4!}\,h(\omega,\pi)\,\varepsilon_{\alpha\beta\lambda\rho}\,\eta^{\alpha}\,\eta^{\beta}\,\eta^{\lambda}\,\eta^{\rho}\;.
$$
The equations of motion $\bar{\partial}G=0$ are then equivalent the appropriate zero rest mass field equations via the direct Penrose transform\footnote{Where $
\phi^{\alpha\beta}(\omega,\pi)=\frac{1}{2}\varepsilon^{\alpha\beta\lambda\rho}\phi_{\lambda\rho}(\omega,\pi)
$.}
$$
H_{AB}(x)=\oint_{S_x}\D^3\pi\;\pi_A\pi_B\,h(\omega,\pi),	\qquad	\Psi_A^{\alpha}(x)=\oint_{S_x}D^3\pi\;\pi_A\,\psi^{\alpha}(\omega,\pi),
$$
\begin{equation}\label{Pen}
\Phi^{\alpha\beta}(x)=\oint_{S_x}\D^3\pi\,\phi^{\alpha\beta}(\omega,\pi)\;.
\end{equation}
The integrals are taken over the $S_x=\C\P^3$ defined by the incidence relation (\ref{Incidence}). 
By virtue of the Penrose transform, these fields satisfy the equations of motion $\nabla^{AC}H_{CB}(x)=0$, $\nabla^{AB}\Psi^{\alpha}_B(x)=0$ and $\Box\Phi^{\alpha\beta}(x)=0$.

\section{The Supersymmetric Penrose Transform}

We wish to construct an integral transform which incorporates each of the bosonic Penrose transforms (\ref{Pen}) in a manifestly supersymmetric way so that we can directly relate the twistor superfield $G({\cal Z})$ to the superspace superfield ${\cal W}^{\alpha\beta}(x,\theta)$. The supersymmetric incidence relations (\ref{Sincidence}) identify a point in complexified superspace with a $\C\P^3$ in supertwistor space with the standard projective measure $\D^3\pi:=\varepsilon^{ABCD}\pi_A\rd \pi_B\rd \pi_C\rd\pi_D$ of projective weight +4. The space-time super field is a projective invariant so, in order to integrate against $D^3\pi$, we need an object of weight -4.  However, the obvious quantity 
$\p^2 G/\p\eta^\alpha\p\eta^\beta$  involves derivatives off the super
quadric \eqref{SQ} into the ambient space and leads to an ill-defined
answer.  Instead we will use $D_A^{\alpha}$, but this does not descend
to twistor space, and so we will have to work on the correspondence
space $ {\cal F}^{9|16}$ with coordinates $(x,\theta,\pi)$ as follows.

The \emph{correspondence space} ${\cal F}^{9|16}\subset{\cal Q}^{6|4} \times {\cal M}^{6|16}$ is defined by the incidence relations \eqref{Sincidence} and inherits from this embedding the double fibration
\begin{align*}
\xymatrix{& {\cal F}^{9|16} \ar[dl]_\mu \ar[dr]^\nu & \\
{\cal Q}^{6|4} & & {\cal M}^{6|16} }
\end{align*}
The fibres of the $\nu$ fibration are $\C\P^{3}$'s and the fibres of
the $\mu$ fibration are super-$\alpha$-planes, locally $\C^{3|12}$. We
can think of ${\cal F}^{9|16}$ as the projective spin bundle over
superspace ${\cal M}^{6|16}$ with coordinates  $(x,\theta,[\pi])$. The
projection $\nu$ is then simply forgetting the $[\pi]$ coordinate, and
the projection $\mu$ is given by mapping to the
$(\omega^A,\pi_A,\eta^\alpha)$ determined by the super incidence relations (\ref{Sincidence}).

On ${\cal F}^{9|16}$ the fibres of $\mu$ are spanned by the operators
\begin{equation}\label{DP}
{\cal D}_{AB}^{\alpha}:=\pi_{[A}\partial^{\alpha}_{B]}+\theta^{\alpha C}\pi_{[A}\nabla_{B]C}\,,	\qquad	\Pi^A:=\pi_B\nabla^{AB}\,.
\end{equation}
Thus a function $G$ pulled back from twistor space is characterized by 
${\cal D}_{AB}^{\alpha}G({\cal Z})=0$, and $\Pi^AG({\cal Z})=0$ so that the correspondence space ${\cal F}^{9|16}$ may be recovered from ${\cal Q}^{6|4} \times {\cal M}^{6|16}$ as the distribution on which $\Pi^A$ and ${\cal D}_{AB}^{\alpha}$ vanish.

\subsection{An integral representation of the direct Penrose Transform}

We now turn to the question of how to construct the correct object of projective weight -4 on $ {\cal F}^{9|16}$ from the lift of the twistor superfield $G({\cal Z})$ from ${\cal Q}^{6|4}$. Once we have this object, we can integrate over the $\C\P^3$ fibres to obtain a superfield on complexified superspace. When acting on functions $G$ on $\cal F$ that are pulled back from
twistor space, so $G=G({\cal Z})$, we can use the incidence relations (\ref{Sincidence}) to write the action of translation and super-translation generators on $G$ in terms of coordinates on supertwistor space. Explicitly we have
$$
\nabla_{AB}G=-\frac{1}{2}\left(\pi_A\frac{\p}{\p \omega^B}-\pi_B\frac{\p}{\p \omega^A}\right)G\,,
$$
and
$$
\frac{\p}{\p\theta^{A\alpha}}G=\left(\pi_A\frac{\p}{\p\eta^{\alpha}}+\frac{1}{2}\Omega_{\alpha\beta}\eta^{\beta}\frac{\p}{\p\omega^A}+\frac{1}{2}\Omega_{\alpha\beta}\pi_A\theta^{B\beta}\frac{\p}{\p\omega^B}\right) G
$$
so that we recover the expression \eqref{DQ} for the supersymmetry generator. The story is a little more complicated for the superspace covariant derivatives, which may be written as
$$
{\cal D}_A^{\alpha}G=-\pi_A\left(\Omega^{\alpha\beta}\frac{\p}{\p\eta^{\beta}}+\theta^{B\alpha}\frac{\p}{\p\omega^B}\right)G\,.
$$
This does not descend to twistor space, as it explicitly depends on superspace coordinate $\theta$; however, ${\cal D}_A^{\alpha}G$ \emph{does} make sense on the product space ${\cal
  Q}^{6|4}\times{\cal M}^{6|16}$ and on the correspondence space ${\cal F}^{9|16}$. When acting on a
twistor function $G$ (a function of $\omega$, $\pi$ and $\eta$ only) we can peel off the $\pi_A$ factor and define a derivative ${\cal D}^{\alpha}G$ such that ${\cal D}^{\alpha}_AG=\pi_{A}{\cal D}^{\alpha}G$ where
$$
{\cal D}^{\alpha}=-\Omega^{\alpha\beta}\frac{\p}{\p\eta^{\beta}}-\theta^{B\alpha}\frac{\p}{\p\omega^B}
$$
and ${\cal D}^{\alpha}G$ is unambiguous if $G$ is a twistor function,
in the sense that the derivative is tangent to $Q^{6|4}$, although
${\cal D}^\alpha G$ no longer lives on twistor space, but on ${\cal
F}^{9|16}$. If we try to define a second derivative ${\cal D}^{\alpha}{\cal
  D}^{\beta}G$ there is no way to peel off the $\pi_A$ factor in
${\cal D}^{\alpha}_A{\cal D}^{\beta}G$ since ${\cal D}^{\alpha}G$ is
not a twistor function as it now depends explicitly on
$\theta^{A\alpha}$. However, we see that 
$$
\pi_{[A}{\cal D}^{\alpha}_{B]}{\cal D}^{\beta}G=
-\Omega^{\alpha\beta}\pi_{[A}\frac{\p}{\p\omega^{B]}}G\, ,
$$
and so, if we consider only the traceless part of ${\cal D}^{\alpha}_A{\cal D}^{\beta}G$, we find
$$
\left(\delta^{\alpha\beta}_{\rho\sigma}+\frac{1}{4}\Omega^{\alpha\beta}\Omega_{\rho\sigma}\right){\cal D}^{\rho}_A{\cal D}^{\sigma}G=\pi_A{\cal D}^{\alpha\beta}G
$$
and the factor of $\pi_A$ can now be removed leading to the definition
of an invariant second order
operator $G\rightarrow {\cal D}^{\alpha\beta}G$.  Explicitly
we have
$$
 {\cal
  D}^{\alpha\beta}:={\cal P}^{\alpha\beta}_{\rho\sigma}\left(\Omega^{\rho\gamma}
\Omega^{\sigma\delta}\frac{\p^2}{\p\eta^{\gamma}\p\eta^{\delta}}
- 2\Omega^{\gamma[\rho}\theta^{\sigma]
  A}\frac{\p^2}{\p\eta^\gamma\p\omega^A}+\theta^{A\rho}\theta^{B\sigma}\frac{\p^2}{\p\omega^A\p\omega^B}\right)\, .
$$
where we have introduced the projections
$$
{\cal P}^{\alpha\beta}_{\rho\sigma}
:=\delta^{\alpha\beta}_{\rho\sigma}+\frac{1}{4}\Omega^{\alpha\beta}\Omega_{\rho\sigma}\,,	\qquad	\widetilde{\cal P}^{\alpha\beta}_{\rho\sigma}:=-\frac{1}{4}\Omega^{\alpha\beta}\Omega_{\rho\sigma}
$$
such that ${\cal P}^{\alpha\beta}_{\rho\sigma}\Omega^{\rho\sigma}=0={\cal P}^{\alpha\beta}_{\rho\sigma}\Omega_{\alpha\beta}$, $\widetilde{\cal P}^{\alpha\beta}_{\rho\sigma}\Omega^{\rho\sigma}=\Omega^{\alpha\beta}$, and $\widetilde{\cal P}^{\alpha\beta}_{\rho\sigma}\Omega_{\alpha\beta}=\Omega_{\rho\sigma}$. The projector ${\cal P}^{\alpha\beta}_{\rho\lambda}$ satisfies the following useful identity
\begin{equation}\label{***}
\Omega_{[\rho|\sigma}{\cal
  P}^{\sigma\tau}_{|\epsilon\lambda]}=\frac{5}{4}\Omega_{[\rho\epsilon}\delta_{\lambda]}^{\tau}\, .
\end{equation}

The object ${\cal D}^{\alpha\beta}G(\cal Z)$ is of projective weight -4 and satisfies the condition $\Omega_{\alpha\beta}{\cal D}^{\alpha\beta}G({\cal Z})=0$ by construction. A candidate for a manifestly supersymmetric integral form of the Penrose transform is therefore
$$
{\cal W}^{\alpha\beta}(x,\theta)=\kappa\oint_{S_x}\D^3\pi\;{\cal D}^{\alpha\beta}G(\omega,\pi,\eta)
$$
where $\kappa$ is a constant. The integrand is defined on the correspondence space ${\cal
  F}^{9|16}\subset{\cal Q}^{6|4}\times{\cal M}^{6|16}$ and integrated
over the $\C\P^3$ fibre $S_x$ over the superspace point $(x,\theta)$
in ${\cal M}^{6|16}$.

We can expand the superfield in $\theta$ and, given the integral
transforms for the bosonic components, obtain the components of the correct
$\theta$ expansion for the superfield ${\cal W}^{\alpha\beta}$. We carry out
this procedure to ${\cal O}(\theta^2)$ in the the Appendix and show that, with the conventions used here, the leading term in the superfield expansion is the scalar field $\Phi^{\alpha\beta}$ if we choose $\kappa=-1$. Our proposal for the (2,0)-supersymmetric direct Penrose transform is thus
\begin{equation}\label{iii}
{\cal W}^{\alpha\beta}(x,\theta)=-\oint_{S_x}\D^3\pi\;{\cal D}^{\alpha\beta}G(\omega,\pi,\eta)
\end{equation}
We have already seen that the scalar superfield ${\cal W}^{\alpha\beta}(x,\theta)$ constructed in this way satisfies the algebraic constraint in (\ref{trace}). In the next section we prove that this integral transform gives a superfield that also satisfies the differential constraint (\ref{constraint}).

\subsection{Superfield constraints}

In order to prove\footnote{The algebraic constraint (\ref{trace}) is automatically satisfied by the inclusion of the projector ${\cal P}^{\alpha\beta}_{\rho\sigma}$ in the definition of ${\cal D}^{\alpha\beta}$.} that the proposed integral transform (\ref{iii}) gives the correct space-time superfield, we must show that the ${\cal W}^{\alpha\beta}$ satisfies the differential constraint (\ref{constraint}). For our purposes, it is useful to write (\ref{constraint}) as
$$
{\cal D}^{\alpha}_A\,{\cal W}^{\beta\gamma}(x,\theta)-\frac{4}{5}\Omega^{\alpha \delta}\;{\cal P}^{\beta\gamma}_{\delta\tau}\,{\cal D}_{A\sigma}{\cal W}^{\sigma \tau}(x,\theta)=0\, ,
$$
and we will show that this constraint naturally follows from \eqref{iii} by taking the derivatives under the integral and showing that the integrand then vanishes.  If we consider now ${\cal P}_{\beta\gamma}^{\rho\sigma}{\cal D}^{\alpha}_A{\cal D}^{\beta}_B{\cal D}^{\gamma}_CG(\omega,\pi,\eta)$ on ${\cal F}$, we observe on the one hand that it is a multiple of $\pi_B\pi_C$ from the discussion of the previous subsection.  
$$
{\cal D}^{\alpha}_A{\cal D}^{\beta}{\cal D}^{\gamma}G(\omega,\pi,\eta)=\left(-\pi_A {\cal D}^{\alpha}{\cal D}^{\beta}{\cal D}^{\gamma} +2\Omega^{[\beta|\tau}\Omega^{\alpha|\gamma]}\frac{\p^2}{\p\eta^{\tau}\p\omega^A}+2\Omega^{\alpha[\gamma}\theta^{\beta]B}\frac{\p^2}{\p\omega^A\p\omega^B}\right)G(\omega,\pi,\eta)
$$
from this we see that
\begin{equation}\label{Z}
{\cal D}^{\alpha}_A{\cal D}^{\rho\sigma}G(\omega,\pi,\eta)={\cal P}^{\rho\sigma}_{\beta\gamma}\left(-\pi_A {\cal D}^{\alpha}{\cal D}^{\beta}{\cal D}^{\gamma} +3\Omega^{[\beta|\tau}\Omega^{|\alpha\gamma]}\frac{\p^2}{\p\eta^{\tau}\p\omega^A}+3\Omega^{[\alpha\gamma}\theta^{\beta]B}\frac{\p^2}{\p\omega^A\p\omega^B}\right)G(\omega,\pi,\eta)
\end{equation}
where we note the fact that the term in parentheses is anti-symmetric in the $\alpha,\beta,\gamma$ indices, since the additional two terms that have been included are projected out by ${\cal P}^{\rho\sigma}_{\beta\gamma}$, hence lifting $\frac{4}{5}\Omega_{\rho\sigma}\Omega^{\alpha \delta}\;{\cal P}^{\beta\gamma}_{\delta\tau}\,{\cal D}^{\rho}_A{\cal W}^{\sigma \tau}$ to ${\cal Q}^{6|4}\times{\cal M}^{9|16}$ gives
$$
\frac{4}{5}\Omega^{\alpha\delta}{\cal P}^{\beta\gamma}_{\delta\tau}\Omega_{[\rho|\sigma}{\cal P}^{\sigma \tau}_{|\epsilon\lambda]}{\cal D}^{\rho}_A{\cal D}^{\epsilon}{\cal D}^{\lambda}G(\omega,\pi,\eta)=\Omega^{\alpha\delta}{\cal P}^{\beta\gamma}_{\delta[\lambda}\Omega_{\rho\epsilon]}{\cal D}^{\rho}_A{\cal D}^\epsilon{\cal D}^{\lambda}G(\omega,\pi,\eta)
$$
where the identity (\ref{***}) has been used to give the expression on the right hand side. Finally, using the fact that ${\cal P}^{\beta\gamma}_{[\delta\lambda}\Omega_{\rho\epsilon]}=0$ we can rewrite ${\cal P}^{\beta\gamma}_{\delta[\lambda}\Omega_{\rho\epsilon]}$ as ${\cal P}^{\beta\gamma}_{[\epsilon\lambda}\Omega_{\rho]\delta}$, so that
$$
\frac{4}{5}\Omega_{\rho\sigma}\Omega^{\alpha\delta}{\cal P}^{\beta\gamma}_{\delta\tau}{\cal D}^{\rho}_A{\cal D}^{\sigma\tau}G(\omega,\pi,\eta)=-{\cal D}^{\alpha}_A{\cal D}^{\beta\gamma}G(\omega,\pi,\eta)
$$
Imposing the incidence relations (\ref{Sincidence}) and integrating over the $\C\P^3$ fibres then gives the supersymmetry constraint equation (\ref{constraint}).

\section{A Superfield for $H^2(n-2)$ representatives}

The direct Penrose transform (\ref{iii}) gives a manifestly supersymmetric map from $H^3$ twistor representatives to components of the space-time superfield. In \cite{Mason:2011nw} a description of a self-dual gerbe, chiral spinor and scalar field - the components of the $(2,0)$ multiplet -  was given in terms of elements of  $H^2(Q;{\cal O}(n-2))$, where $n/2$ is the spin of the field in question. The  $H^2(Q;{\cal O}(n-2))$ fields are related to their  $H^3(Q;{\cal O}(-n-4))$ counterparts by \cite{Mason:2011nw}

$$
\bar{\p}b(\omega,\pi)=\frac{1}{3!}(\omega\cdot\pi)^3h(\omega,\pi)\;,	\qquad	\bar{\p}\chi^{\alpha}(\omega,\pi)=-\frac{1}{2!}(\omega\cdot\pi)^2\psi^{\alpha}(\omega,\pi)\;,
$$
\begin{equation}\label{fn}
\bar{\p}\varphi^{\alpha\beta}(\omega,\pi)=(\omega\cdot\pi)\phi^{\alpha\beta}(\omega,\pi)\;,
\end{equation}
where $\{h,\psi,\phi\}\in H^3(Q;{\cal O}(-n-4))$ and $\{b,\chi,\varphi\}\in H^2(Q;{\cal O}(n-2))$ and convenient constants of proportionality have been chosen. In addition to the superfield $G(\omega,\pi,\eta)$ already considered, we may introduce a superfield $C(\omega,\pi,\eta)$ of projective weight zero,\begin{equation}\label{C}
C(\omega,\pi,\eta)=b(\omega,\pi)+\chi_{\alpha}(\omega,\pi)\,\eta^{\alpha}+\frac{1}{2}\varphi_{\alpha\beta}(\omega,\pi)\,\eta^{\alpha}\,\eta^{\beta} \;,
\end{equation}
the bosonic components of which give the $H^2(Q,{\cal O}(n-2))$ description of the linear $(2,0)$ theory. Note that we could, in principle, introduce terms at order $\eta^3$ and $\eta^4$. Indeed, such terms would be required for off-shell closure of the supersymmetry algebra; however, most of our considerations here will be on-shell and we do not consider these extra redundant terms.

We can derive the space-time  superfields from $C({\cal Z})$ directly via an indirect Penrose transform. This transform, which does not have a straightforward integral representation in six dimensions, gives a space-time potential for the superfields. 

A potential for the  ${\cal H}_{AB}(x,\theta)$ superfield arises from $C({\cal Z})$ by an argument similar to that for recovering the self-dual gerbe $H_{AB}(x)$ from the twistor representative $b(Z)\in H^2(Q';{\cal O})$ (see the Appendix of \cite{Mason:2011nw} and also \cite{Saemann:2011nb}) which we restrict a point on superspace, by choosing a $\C\P^3\in{\cal Q}$. Using similar arguments, we shall propose a relation between the twistor superfield $C({\cal Z})$ and the spacetime superfield ${\cal W}^{\alpha\beta}$.

Let $f(Z)$ denote an element from the set of components $\{b(Z),\chi_{\alpha}(Z),\varphi_{\alpha\beta}(Z)\}$ of the superfield $C(\cal Z)$ given by (\ref{C}). Following \cite{Chatterjee} the components $f$ can be understood via \v Cech cohomology: let $[f]$ be a representative of $\check{H}^2(n-2)$ and $\{U_i\}$ a Leray cover\footnote{A Leray cover is one for which the open sets have no cohomology so that the \v Cech cohomology agrees with the standard cohomology.} of $Q$. We then have a family of functions of homogeneity degree $n-2$, $[f]=\{f_{ijk}\}$ defined on the triple intersection
$$
f_{ijk}:U_i\cap U_j\cap U_k\rightarrow \C^* \, , \quad f_{ijk} f_{jkl}f_{kli}f_{lij}=1.
$$
Restricting to $S_x=\C\P^3$, $\check{H}^2(n-2)=0$ for $n=0,1,2$ and so we can write
$$
f_{ijk}=a_{ij}a_{jk}a_{ki} \, .
$$
We can do this for each component of the superfield (\ref{C}) so that, on restriction to $S_x$, we have
\begin{equation}\label{split1}
\log C_{ijk}=\log A_{ij}+\log A_{jk}+\log A_{ki} \, ,
\end{equation}
where $\log A_{ij}$ includes contributions from each of the components $a_{ij}$ restricted to $U_i\cap U_j$. We note that $C_{ijk}$ is defined on ${\cal Q}^{6|4}$ and may be lifted to the correspondence space ${\cal F}^{9|16}=\{(x,\theta,[\pi])\in\C^{6|16}\times \C\P^3\}$ where, by virtue of being pulled back from supertwistor space, it satisfies
$$
\mu^*\left(\Pi^AC_{ijk}\right)=0 \;,	\qquad	\mu^*\left(D_{AB}^{\alpha}C_{ijk}\right)=0\;.
$$
If we think of ${\cal F}^{9|16}$ as a double fibration, $\Pi^A$ and $D^{\alpha}_{AB}$ are differential operators along the fibres of the $\mu$ fibres (\ref{DP}). In contrast, $A_{ij}=A_{ij}(x,\pi,\theta)$ is not pulled back from supertwistor space\footnote{Assuming the components $[f]$ were not trivial.} and so
$\mu^*\left(\Pi^AA_{ij}\right)$ and $\mu^*\left(D_{AB}^{\alpha}A_{ij}\right)$ are not zero. It is useful to define $A_{ijAB}{}^{\alpha}=D_{AB}^{\alpha}\log A_{ij}$, and differentiating \eqref{split1} we obtain
$$
A_{ijAB}{}^{\alpha}+A_{jkAB}{}^{\alpha}+A_{kiAB}{}^{\alpha}=0 \, .
$$
It is also the case \cite{Baston:1989vh} that $\check{H}^1(S_x;{\cal O^*})=0$ and so there exist $\lambda_{iAB}^{\alpha}$ such that
\begin{equation}\label{split2}
A_{ijAB}{}^{\alpha}=\lambda_{iAB}^{\alpha}-\lambda_{jAB}^{\alpha}\, .
\end{equation}
Furthermore, we can use the fact that
\begin{equation}\label{PDD}
{\cal P}^{\alpha\beta}_{\rho\lambda}\{D_{AB}^{\rho},D_{CD}^{\lambda}\}=0
\end{equation}
to differentiate \eqref{split2} and obtain
$$
{\cal P}^{\alpha\beta}_{\rho\lambda}(D^{\lambda}_{AB}\lambda_{iCD}^{\rho}-D^{\lambda}_{CD}\lambda_{iAB}^{\rho})={\cal P}^{\alpha\beta}_{\rho\lambda}(D^{\lambda}_{AB}\lambda_{jCD}^{\rho}-D^{\lambda}_{CD}\lambda_{jAB}^{\rho}):=S^{\alpha\beta}_{AB:CD}\;.
$$
The left and right hand sides of the first equality are defined on different patches ($U_i$ and $U_j$) but are equal. From this we infer the existence of the globally defined field $S^{\alpha\beta}_{AB:CD}$ which is homogenous of degree two in $\pi_A$. $S^{\alpha\beta}_{AB:CD}$ is antisymmetric in each pair of indices $[AB]$ and $[CD]$ and is also pairwise antisymmetric. We can express the $\pi$ dependence in $S^{\alpha\beta}_{AB:CD}$ explicitly as
$$
S^{\alpha\beta}_{AB:CD}=\pi_A\pi_CB^{\alpha\beta}_{BD}-\pi_B\pi_CB^{\alpha\beta}_{AD}+\pi_B\pi_DB^{\alpha\beta}_{AC}-\pi_A\pi_DB^{\alpha\beta}_{BC}\,,
$$
in terms of a space-time potential $B^{\alpha\beta}_{AB}=-B^{\alpha\beta}_{BA}$, which satisfies $\Omega_{\alpha\beta}B^{\alpha\beta}_{AB}=0$. This field is a potential for the scalar superfield
$$
{\cal W}^{\alpha\beta}(x,\theta)=\nabla^{AB}B^{\alpha\beta}_{AB}(x,\theta)
$$
Similar constructions can be used to find potentials $B^{\alpha}_{ABC}$ and $B^A{}_B$ using $[\Pi^A,D_{BC}^{\alpha}]=0$ and $[\Pi^A,\Pi^B]=0$ respectively, in place of (\ref{PDD}). These potentials are then related to spin-half and spin-one superfields as $\Xi^{\alpha}_A\sim \nabla^{BC}B^{\alpha}_{ABC}$ and ${\cal H}_{AB}\sim\nabla_{(A|C}B^C{}_{|B)}$ with appropriate normalisations.

We expect the $C(\cal Z)$ superfield to describe the same space-time physics as the $G(\cal Z)$ superfield. The two can clearly be related by composing direct and indirect Penrose transforms; however, we can find a more direct relationship between the superfields if we consider the obstructions to extending $C(\cal Z)$ off the superquadric ${\cal Q}$. The notion of extending a field off the quadric can be made precise by considering formal neighbourhoods of the quadric \cite{Griffiths,Witten:1978xx,Eastwood:1986,Pool}, that is by introducing an auxiliary parameter $\xi$, which acts as a coordinate along a direction off the quadric and into the ambient $\C\P^{7|4}$. We then allow the field to depend on this coordinate in a limited way. In particular, an extension of the field into the $N$'th formal neighbourhood of the quadric allows the field to have a polynomial dependence on $\xi$ up to and including order $N$. To enforce this we require $\xi^{N+1}=0$, so it is particularly useful to think of $\xi$ as a Grassmann coordinate. In the case at hand, we consider the superfield $G$ on the second formal neighbourhood of ${\cal Q}^{6|4}$, were we introduce additional grassmann coordinates $\xi$ such that
$$
{\cal Z}^2=\xi^2		\qquad	\text{and}	\qquad	\xi^n=0	\qquad \text{where $n>3$}	
$$
then
$$
\omega\cdot\pi=\xi^2+\frac{1}{2}\Omega_{\alpha\beta}\eta^{\alpha}\eta^{\beta}	\,,	\qquad	(\omega\cdot\pi)^2=\xi^2\,\Omega_{\alpha\beta}\eta^{\alpha}\eta^{\beta}+\frac{1}{12}\varepsilon_{\alpha\beta\lambda\rho}\eta^{\alpha}\eta^{\beta}\eta^{\lambda}\eta^{\rho}\,,
$$
$$
(\omega\cdot\pi)^3=\frac{1}{4}\xi^2\,\varepsilon_{\alpha\beta\lambda\rho}\eta^{\alpha}\eta^{\beta}\eta^{\lambda}\eta^{\rho}\,,
$$
and $(\omega\cdot\pi)^n=0$ for $n>3$.

Given the formal neighbourhood relations (\ref{fn}) we can evaluate $\bar{\p}C$ on the second formal neighbourhood of ${\cal Q}^{6|4}$
$$
\bar{\p}C=\frac{1}{6}(\omega\cdot\pi)^3h-\frac{1}{2}(\omega\cdot\pi)^2\Omega_{\alpha\beta}\psi_{\beta}\eta^{\alpha}+\frac{1}{2}(\omega\cdot\pi)\phi_{\alpha\beta}\,\eta^{\alpha}\,\eta^{\beta}\;,
$$
then, using $\varepsilon_{\alpha\beta\lambda\rho}=3\Omega_{[\alpha\beta}\Omega_{\lambda\rho]}$ and $\phi_{[\alpha\beta}\Omega_{\lambda\rho]}\sim\varepsilon_{\alpha\beta\lambda\rho}(\Omega\cdot\phi)=0$, the above expression for $\bar{\p}C$ may be written in terms of the superfield expansion in powers of $\eta$
$$
\bar{\p}C=\xi^2\left(\frac{1}{2}\phi_{\alpha\beta}(Z)\;\eta^{\alpha}\,\eta^{\beta}+\frac{1}{6}\,\psi^{\alpha}(Z)\;\varepsilon_{\alpha\beta\gamma\delta}\,\eta^{\beta}\,\eta^{\gamma}\,\eta^{\delta}+\frac{1}{24}\,h(Z)\;\varepsilon_{\alpha\beta\gamma\delta}\,\eta^{\alpha}\,\eta^{\beta}\,\eta^{\gamma}\,\eta^{\delta}\right)\;.
$$
We recognise the expression in parenthesis as the $\eta$ expansion of $G(\cal Z)$. Thus, on the second formal neighbourhood, the superfields $G$ and $C$ are related by
\begin{equation}\label{sfn}
\bar{\p}C={\cal Z}^2 G\,.
\end{equation}
This is the supersymmetric analogue of the bosonic relations (\ref{fn}).

\section{Supertwistor actions}

As a first step towards constructing twistor actions  for linear chiral supersymmetric theories, we consider the problem of finding an action for the linear self-dual gerbe $H_{AB}(x)$. This field is described in twistor space by either the representatives of either $H^2(Q;{\cal O})$ or $H^3(Q;{\cal O}(-6))$, via the indirect and direct Penrose transforms respectively.  To construct twistor actions it is most convenient to use Dolbeault representatives of the cohomology classes $H^2(n-2)$ and $H^3(-n-4)$. Introducing the $(0,2)$- and $(0,3)$-forms $b(Z)$ and $h(Z)$ respectively, we can write an action in twistor space for the linear self-dual gerbe
$$
S=\int_{\C\P^7} \D^7Z\wedge\bar{\delta}(Z^2)\wedge h\wedge\bar{\partial}\,b\;,
$$
where $\bar{\delta}(Z^2)=\bar{\p}(Z^{-2})$ is a $(0,1)$-form of projective weight -2 and $\D^7 Z=\varepsilon_{\alpha_0\alpha_2....\alpha_7}Z^{\alpha_0}\p Z^{\alpha_1}\wedge...\wedge\p Z^{\alpha_7}=\D^3\pi\rd^4 \omega$ is the natural projective $(7,0)$ form on $\C\P^7$ of weight +8. $h$ is a $(0,3)$-form of weight -6 and $b$ is a $(0,2)$-form of weight 0, so this action is well-defined. The action has the gauge symmetry
$$
b\rightarrow b+\bar{\p}\lambda\;,	\qquad	h\rightarrow h+\bar{\p}\Lambda\;,
$$
where $\lambda(Z)$ and $\Lambda(Z)$ are arbitrary $(0,1)$- and $(0,2)$-forms respectively. Varying the action with respect to either $b$ or $h$ gives the condition that $b$ and $h$ are holomorphic which, in addition to the gauge symmetry, ensures that $b\in H^2(Q;{\cal O})$ or $h\in H^3(Q;{\cal O}(-6))$, and so, by virtue of the Penrose transform (\ref{Pen}), describe an on-shell self-dual gerbe.

We can pick out the twistor space for real (compactified) Minkowski space in $\M$ as the sub-space of $Q$ for which $Z\cdot\hat{Z}=0$, where $\hat{Z}$ denotes quaternionic conjugation on the spinor components. An action corresponding to a linear self-dual gerbe theory on Minkowski space is then
\begin{equation}\label{Mink}
S=\int_{\C\P^7|_{Z\cdot\hat Z=0}} \D^7Z\wedge\bar{\delta}(Z^2) \wedge t\wedge\bar{\partial}\,b\;,
\end{equation}
for some twistor field $t$. The equations on motion imply that $b\in H^2(Q;\cO)$ as it should and $t\in H^2(Q;\cO(-4))$; however, $H^2(Q;\cO(-4))=0$ and so $t$ has no on-shell degrees of freedom and acts simply as a Lagrange multiplier, constraining $b$ to lie in $H^2(Q;\cO)$.

Let us now consider the supersymmetric extension. We consider general $(0,2)$- and $(0,3)$-form superfields $C(\omega,\pi,\eta)$ and $G(\omega,\pi,\eta)$ respectively. A natural supersymmetric action is
\begin{equation}\label{action}
S=\int_Y\D^{7|4}{\cal Z}\wedge\bar{\delta}({\cal Z}^2)\wedge G\wedge\bar{\partial}\, C\;,
\end{equation}
where $Y\subset \C\P^{7|4}$ is a $(14|4)$ (real) dimensional space such that $\bar{\eta}^{\alpha}=0$ and $\D^{7|4}{\cal Z}=\D^3\pi\rd^4\omega\rd^4\eta$. The action is invariant under the gauge symmetries 
$$
C\rightarrow C+\bar{\p}\lambda\;,	\qquad	G\rightarrow G+\bar{\p}\Lambda\;,
$$
where $\lambda$ and $\Lambda$ are now superfields. The equation of motion for $G$ arising from this action are
$$
\bar{\partial}G=0,	\qquad	\bar{\p}C=0\;,
$$
so that, if we expand $G$ and $C$ in powers of $\eta$, the bosonic coefficients take values in $H^3(Q,{\cal O}(-n-4))$ and $H^2(Q,{\cal O}(n-2))$ respectively.

\subsection{(4,0) Superconformal Theory}

Up until now we have been concerned with spin-one gerbes, here we extend our considerations to spin-two fields. The on-shell graviton is given by the field strength $\Psi^{ab}_{\dot{a}\dot{b}}$ which has six degrees of freedom, as we would expect; however, the spin-two field strength $G_{abcd}$ arising from the direct Penrose transform
\begin{equation}\label{G}
G_{ABCD}=\oint_{S_x}\D^3\pi\,\pi_A\pi_B\pi_C\pi_D\;g(\omega,\pi)
\end{equation}
has five on-shell degrees of freedom. Furthermore $G_{ABCD}$ is a \emph{chiral} field, whereas the graviton is not. This spin-two field appearing from twistor space is clearly not describing linearised Einstein gravity, but a more exotic six-dimensional relative. It is conjectured that there exists a superconformal $(4,0)$ theory in six-dimensions \cite{Hull:2000ih,Hull:2000rr,Hull:2000zn,Schwarz:2000zg} which includes just such a field. The novelty of this theory is that the spin-two field is not a graviton and is not thought to give rise to a conventional, geometric, theory of gravitation. Rather, the spin-two field is given by a tensor $C_{\mu\nu\lambda\rho}$ with the symmetries of the Riemann tensor and field strength
$$
G_{\mu\nu\lambda\rho\sigma\eta}=3\partial_{\mu}\partial_{[\nu}C_{\lambda\rho]\sigma\eta}+3\partial_{\eta}\partial_{[\nu}C_{\lambda\rho]\mu\sigma}+3\partial_{\sigma}\partial_{[\nu}C_{\lambda\rho]\eta\mu}
$$
which is self-dual
$$
G_{\mu\nu\lambda abc}=\varepsilon_{abcdef}G_{\mu\nu\lambda}{}^{def}
$$
In terms of spinor notation, the five physical degrees of freedom are given by a completely symmetric field $C_{AB}{}^{CD}$ with (linearised) manifestly self-dual field strength
$$
G_{ABCD}=\nabla_{(A|M}\nabla_{|B|N}C_{|CD)}{}^{MN}
$$
This field, with 5 on-shell `gravi-gerbe' degrees of freedom, is the highest spin member of the $(4,0)$ multiplet, which includes; 32 `non-geometric gravigerbini' $\Psi^{\alpha}_{ABC}$, 81 self-dual gerbes $H_{AB}^{\alpha\beta}$, 96 gerbini $\lambda^{\alpha\beta\rho}_A$, and 42 scalars $\Phi^{\alpha\beta\rho\sigma}$. $\alpha,\beta,\rho,\sigma=1,2,..8$ are $SU(8)$ R-symmetry indices. At the linearised level, a dimensional reduction on a circle to five dimensions yields the linearised form of the conventional Einstein maximal supergravity in five dimensions and it is conjectured that there exists a non-linear $(4,0)$ theory in six-dimensions which gives rise to the full Einstein supergravity in five dimensions \cite{Hull:2000zn}. It is not clear what the full non-linear $(4,0)$ theory should look like but it is expected that the interactions will not be of a conventional field-theoretic type but rather should be based on yet to be identified M-Theoretic principles. In this section we consider only the linearised form of the $(4,0)$ theory in supertwistor space.

The spin two field may be described in terms of the conventional Penrose transform (\ref{G}) where $g(\omega,\pi)\in H^3(Q;{\cal O}(-8))$ and it is straightforward to show that a representative of $H^2(Q;{\cal O}(+2))$ will give a description of this field in terms of a potential $C_{ABCD}$, modulo gauge-invariance, from the indirect Penrose transform. Similar expressions exist for the other fields in the $(4,0)$ multiplet. A natural on-shell superfield for the $(4,0)$ multiplet is\footnote{As in the $(2,0)$ case, the off-shell superfield appearing in the action will have extra terms required by the off-shell closure of the superalgebra.}
$$
\G=g\,(\eta^8)+\psi^{\alpha}\,(\eta^7)_{\alpha}+h^{\alpha\beta}\,(\eta^6)_{\alpha\beta}+\lambda^{\alpha\beta\lambda}\,(\eta^5)_{\alpha\beta\lambda}+\phi^{\alpha\beta\lambda\rho}\,(\eta^4)_{\alpha\beta\lambda\rho}\;,
$$
where
$$
\Omega_{\alpha\beta}h^{\alpha\beta}=0\;,	\qquad	\Omega_{\alpha\beta}\lambda^{\alpha\beta\lambda}=0\;,	\qquad	\Omega_{\alpha\beta}\Phi^{\alpha\beta\lambda\rho}=0\;,
$$
and $\Omega_{\alpha\beta}=-\Omega_{\alpha\beta}$ is now the invariant of the $R$-symmetry group $SU(8)$. These conditions impose one, eight and twenty-eight constraints on $h^{\alpha\beta}$, $\lambda^{\alpha\beta\rho}$ and $\Phi^{\alpha\beta\rho\sigma}$ respectively, giving the correct number of on-shell degrees of freedom once we perform a Penrose transform on these twistor functions. We can also consider an $H^2(Q,{\cal O}(n-2))$ description of the super multiplet in terms of a twistor superfield of projective weight +2 
$$
{\cal C}=C+\Psi_{\alpha}\,\eta^{\alpha}+\frac{1}{2} b_{\alpha\beta}\,\eta^{\alpha}\eta^{\beta}+\frac{1}{3!}\lambda_{\alpha\beta\lambda}\,\eta^{\alpha}\eta^{\beta}\eta^{\lambda}+\frac{1}{4!}\Phi_{\alpha\beta\lambda\rho}\,\eta^{\alpha}\eta^{\beta}\eta^{\lambda}\eta^{\rho}\;,
$$
with similar constraints on the bosonic field components. Actions akin to those given for the $(2,0)$ case may be constructed 
$$
S=\int_Y \D{\cal Z}^{7|8}\wedge\bar{\delta}({\cal Z}^2)\wedge \G\wedge\bar{\partial}{\cal C}\;,
$$
and one expects a condition, from the second formal neighbourhood
$$
\dbar {\cal C}={\cal Z}^2\G
$$
to hold in this case also, where now ${\cal C}({\cal Z})$ is of homogeneity +2 and $\G({\cal Z})$ is of homogeneity zero.

\section{Discussion}

We have provided a description, in terms of supertwistor geometry of linearised conformal field theories in six-dimensions. Our main motivation was to gain some insight on the (2,0) conformal theory; however, the ideas presented here also generalise to the (4,0), maximally supersymmetric theory. In principle, it should not be too hard to generalise the integral Penrose transform (\ref{iii}) to the (4,0) case to find a space-time superfield
$$
{\cal R}^{\alpha\beta\lambda\rho}(x,\theta)=\kappa\oint_{S_x}\D^3\pi\;{\cal D}^{\alpha\beta\lambda\rho}{\cal G}(\omega,\pi,\eta)\;,
$$
which encodes the fields of the (4,0) multiplet, where ${\cal D}^{\alpha\beta\lambda\rho}\sim {\cal P}^{\alpha\beta\lambda\rho}_{\sigma\delta\epsilon\eta}{\cal D}^{\sigma}{\cal D}^{\delta}{\cal D}^{\epsilon}{\cal D}^{\rho}$ and ${\cal P}^{\alpha\beta\lambda\rho}_{\sigma\delta\epsilon\eta}$ is a projector than ensures that any contraction of ${\cal R}^{\alpha\beta\lambda\rho}$ with the the R-symmetry invariant tensor $\Omega$ will vanish and ${\cal D}^{\alpha}_A{\cal G}=\pi_A{\cal D}^{\alpha}{\cal G}$
if ${\cal G}$ is a supertwistor field where ${\cal D}^{\alpha}_A$ is (4,0) supercovariant derivative. In principle, the differential constraint that such an ${\cal R}^{\alpha\beta\lambda\rho}(x,\theta)$ should satisfy should follow from this construction; however, we have not investigated this possibility.

The constructions considered in this article have an obvious formal generalisation to $(2N,0)$ \emph{linearised} theories (interacting theories with massless fields of spin higher than two are problematic), where the highest spin field has spin $n$ and field strength given by
$$
F_{A_1A_2...A_{2N}}=\oint_{S_x}\D^3\pi\,\pi_{A_1}\pi_{A_2}...\pi_{A_{2N}}\,f(\omega,\pi)\;.
$$
where $f(\omega,\pi)\in H^3(Q;{\cal O}(-n-4))$. Alternatively the field may be described by a potential, modulo gauge, as $f\in H^2(Q;{\cal O}(n-2))$. The action for the linearised theory in this general case is the same as that for the $(2,0)$ and $(4,0)$ theories; however, any speculative interaction terms one might consider will differ for each $N$. One may formally construct a superfield of homogeneity $2N-4$
$$
{\cal F}=\sum_{k=0}^{4N-2}f^{\alpha_1\alpha_2...\alpha_{k}}\,(\eta^{4N-k})_{\alpha_1\alpha_2...\alpha_k}\;,	\qquad	\text{where}	\qquad	(\eta^{4N-k})_{\alpha_1\alpha_2...\alpha_k}:=\frac{1}{k!}\varepsilon_{\alpha_1...\alpha_k...\alpha_{4n}}\eta^{\alpha_{k+1}}...\eta^{\alpha_{4N}}
$$
We can also consider a weight $2N-2$ superfield
$$
{\cal A}=\sum_{k=0}^N\frac{1}{k!}A_{\alpha_1...\alpha_{k}}\eta^{\alpha_1}...\eta^{\alpha_{k}}\;,
$$
and a topological action
$$
S=\int \D{\cal Z}^{7|2N}\wedge\bar{\delta}({\cal Z}^2)\wedge {\cal F}\wedge\dbar {\cal A}\;,
$$
where 
$$
\D{\cal Z}^{7|2N}=\D^3\pi\rd^4\omega\rd\eta^{2N}\;.
$$
and whose equation of motion ensures that ${\cal A}$ and ${\cal F}$ are holomorphic. Many of the results given in the previous sections for the $(2,0)$ theory generalise to $(2N,0)$ supersymmetry. It would be interesting to see if these ideas can be usefully applied to study higher spin field theory. In particular, one might expect such ideas to play a role in any proposed unbroken superconformal phase of M-Theory.  Finally, we should express the hope that the considerations presented here might provide a useful framework in constructing \emph{interacting} (2,0) and possibly even (4,0) conformal theories \cite{Witten:1995zh,Seiberg:1996vs,Witten:1996hc}. In this article we have restricted our considerations to field theories. If these ideas are to have application to understanding interacting theories, one may be forced to consider a generalisation of the known twistor string theories \cite{Witten:2003nn,Berkovits:2004hg,Mason:2007zv} to have any hope of describing say the dynamics, say of little string theories \cite{Seiberg:1997zk,Strominger:1995ac,Aharony:1999ks}, in a twistorial setting.

\begin{center}
\textbf{Acknowledgements}
\end{center}

We would like to thank Martin Wolf for discussions. LM is supported by a Leverhulme Fellowship.

\appendix

\section{Recovering the component form of the space-time superfield}

That ${\cal W}^{\alpha\beta}$ satisfies the constraints (\ref{trace}) and (\ref{constraint}) is sufficient to identify ${\cal W}^{\alpha\beta}$ as the $(2,0)$ superfield; however, It is still informative to see precisely how the complicated superspace component expansion arises naturally from the very simple integral formula for the Penrose transform (\ref{iii}). In particular, we find that the superquadric geometry of ${\cal Q}^{6|4}$ explains why the space-time description of the six-dimensional superfield is so much more complicated than that of the ${\cal N}=4$ superfield in four dimensions, even though the component superfields on supertwistor space are of comparable complexity. We start with the $(0,3)$-form superfield
$$
G(\omega,\pi,\eta)=\frac{1}{2}\phi_{\alpha\beta}(\omega,\pi)\;\eta^{\alpha}\,\eta^{\beta}+\frac{1}{6}\,\psi^{\alpha}(\omega,\pi)\;\varepsilon_{\alpha\beta\gamma\delta}\,\eta^{\beta}\,\eta^{\gamma}\,\eta^{\delta}+\frac{1}{24}\,h(\omega,\pi)\;\varepsilon_{\alpha\beta\gamma\delta}\,\eta^{\alpha}\,\eta^{\beta}\,\eta^{\gamma}\,\eta^{\delta}
$$
We shall only consider the superfield to order $\theta^2$; however, the general procedure for determining the full superfield expansion will be made clear. The relevant terms arise from
$$
-{\cal D}^{\alpha\beta}G(\omega,\pi,\eta)=-{\cal P}^{\alpha\beta}_{\lambda\rho}\left(\Omega^{\lambda\sigma}\Omega^{\rho\gamma}\frac{\p^2}{\p\eta^{\sigma}\p\eta^{\gamma}}G(\omega,\pi,\eta)-2\Omega^{\lambda\sigma}\theta^{A\rho}\frac{\p^2}{\p\omega^A\p\eta^{\sigma}}\left(\frac{1}{2}\phi_{ij}(\omega,\pi)\eta^i\eta^j+...\right)+...\right)
$$
where $+...$ denote terms that give rise to ${\cal O}(\theta^3)$ terms in the space-time superfield. Expanding this expression out
\begin{eqnarray}
-{\cal D}^{\alpha\beta}G(\omega,\pi,\eta)&=&{\cal P}^{\alpha\beta}_{\lambda\rho}\left(\phi^{\lambda\rho}+2\left(\delta^{\lambda\rho}_{ij}+\frac{1}{2}\Omega^{\lambda\rho}\Omega_{ij}\right)\psi^i(\omega,\pi)\eta^j+\left(\delta^{\lambda\rho}_{ij}+\frac{1}{2}\Omega^{\lambda\rho}\Omega_{ij}\right)\,h(\omega,\pi)\,\eta^i\eta^j\right.\nonumber\\
&&\left.-2\Omega^{\lambda\sigma}\theta^{A\rho}\p_A\,\phi_{\sigma l}(\omega,\pi)\,\eta^l+...\right)\;.\nonumber
\end{eqnarray}
Using the properties of ${\cal P}^{\alpha\beta}_{\lambda\rho}$, this simplifies to
$$
-{\cal D}^{\alpha\beta}G(\omega,\pi,\eta)={\cal P}^{\alpha\beta}_{\lambda\rho}\left(\phi^{\lambda\rho}(\omega,\pi)+2\psi^{\lambda}(\omega,\pi)\eta^{\rho}+h(\omega,\pi)\,\eta^{\lambda}\eta^{\rho}-2\Omega^{\lambda\sigma}\theta^{A\rho}\p_A\phi_{\sigma l}(\omega,\pi)\eta^l+...\right)\;.
$$
Each of the component fields $\phi^{\alpha\beta}(\omega,\pi)$, $\psi^{\alpha}(\omega,\pi)$ and $h(\omega,\pi)$ depend on $\theta^{A\alpha}$ via the incidence relations (\ref{Sincidence}). The full superfield ${\cal W}^{\alpha\beta}(x,\theta)$ is given by imposing the incidence relations on $-{\cal D}^{\alpha\beta}G(\omega,\pi,\eta)$ so that all explicit $\omega$- and $\eta$-dependences are removed, expanding about $\theta^{A\alpha}=0$ and finally integrating over the $\C\P^3$ fibres of ${\cal F}^{9|16}$.

Imposing the incidence relations (\ref{Sincidence}) and expanding in powers of $\theta$, using
$$
\phi^{\alpha\beta}(\omega,\pi)=\phi^{\alpha\beta}(x,\pi)-\frac{1}{2}\Omega_{\rho\lambda}\theta^{A\rho}\theta^{B\lambda}\pi_{(A}\p_{B)}\phi^{\alpha\beta}(x,\pi)+...\,,
$$
and keeping only terms of order $\theta^2$, gives
\begin{eqnarray}
-{\cal D}^{\alpha\beta}G(\omega,\pi,\eta)&=&{\cal P}^{\alpha\beta}_{\lambda\rho}\left(\phi^{\lambda\rho}(x,\pi)+2\psi^{\lambda}(x,\pi)\pi_A\theta^{A\rho}+h(x,\pi)\,\pi_A\pi_B\theta^{A\lambda}\theta^{B\rho}\right.
\nonumber\\
&&\left.-2\Omega^{[\lambda|\sigma}\theta^{A|\rho]}\theta^{Bl}\pi_B\p_A\phi_{\sigma l}(x,\pi)-\frac{1}{2}\Omega_{ij}\theta^{Ai}\theta^{Bj}\pi_{(A}\p_{B)}\phi^{\lambda\rho}(x,\pi)+...\right)\;,\nonumber
\end{eqnarray}
where now each of the bosonic components is a function of $(x,\pi)$. We can write the derivative terms above as
$$
-{\cal P}^{\alpha\beta}_{\lambda\rho}\Omega^{\lambda\sigma}\Omega^{\rho k}\theta^{Ai}\theta^{Bj}\left(2\Omega_{ki}\pi_B\p_A\phi_{\sigma j}(x,\pi)+\frac{1}{2}\Omega_{ij}\pi_{(A}\p_{B)}\phi_{\sigma k}(x,\pi)\right)\;,
$$
which can be written as
\begin{eqnarray}\label{m}
&&-\frac{1}{2}{\cal P}^{\alpha\beta}_{\lambda\rho}\Omega^{\lambda\sigma}\Omega^{\rho k}\theta^{Ai}\theta^{Bj}\Big(\Omega_{ki}\pi_B\p_A\phi_{\sigma j}(x,\pi)-\Omega_{\sigma i}\pi_B\p_A\phi_{k j}(x,\pi)-\Omega_{kj}\pi_A\p_B\phi_{\sigma i}(x,\pi)
\nonumber\\
&&+\Omega_{\sigma j}\pi_A\p_B\phi_{k i}(x,\pi)+\Omega_{ij}\pi_{A}\p_{B}\phi_{\sigma k}(x,\pi)\Big)\;.
\end{eqnarray}
Using the fact that $\Omega_{\alpha\beta}\phi^{\alpha\beta}=0$ and $\Omega_{[kj}\phi_{\sigma i]}\sim\varepsilon_{kj\sigma i}\Omega_{\alpha\beta}\phi^{\alpha\beta}$ we are lead to the six-term identity
$$
\Omega_{kj}\phi_{\sigma i}+\Omega_{ki}\phi_{j\sigma}+\Omega_{k \sigma}\phi_{ij}-\Omega_{ij}\phi_{\sigma k}-\Omega_{\sigma i}\phi_{jk}-\Omega_{j\sigma}\phi_{ik}=0\;,
$$
which can be used to re-write the last three terms in (\ref{m}) to give
\begin{equation}\label{n}
-\frac{1}{2}{\cal P}^{\alpha\beta}_{\lambda\rho}\Omega^{\lambda\sigma}\Omega^{\rho k}\theta^{Ai}\theta^{Bj}\Big(4\Omega_{\sigma i}\pi_{[A}\p_{B]}\phi_{k j}(x,\pi)+\Omega_{k\sigma}\pi_{A}\p_{B}\phi_{ij}(x,\pi)\Big)\;.
\end{equation}
The last term of (\ref{n}) is pure trace and vanishes upon contraction with ${\cal P}^{\alpha\beta}_{\lambda\rho}$, leaving
\begin{eqnarray}
-{\cal D}^{\alpha\beta}G(\omega,\pi,\eta)&=&{\cal P}^{\alpha\beta}_{\lambda\rho}\left(\phi^{\lambda\rho}(x,\pi)
+2\pi_A\;\psi^{\lambda}(x,\pi)\theta^{A\rho}+\pi_A\pi_B\;h(x,\pi)\,\theta^{A\lambda}\theta^{B\rho}\right.
\nonumber\\
&&-\left.2\theta^A_{\sigma}\theta^B_l\Omega^{[\lambda|\sigma}\pi_{[A}\p_{B]}\phi^{|\rho]l}(x,\pi)+...\right)\;.
\end{eqnarray}
Integrating over $\C\P^3$ fibres of the $\nu$ fibration with homogenous coordinates $\pi_A$ and using the Penrose transforms (\ref{Pen}), we have (\ref{iii})
$$
{\cal W}^{\alpha\beta}(x,\theta)={\cal P}^{\alpha\beta}_{\lambda\rho}\left(\Phi^{\lambda\rho}(x)+2\Psi_A^{\lambda}(x)\theta^{A\rho}+H_{AB}(x)\,\theta^{A\lambda}\theta^{B\rho}+2\theta^A_{\sigma}\theta^B_l\Omega^{[\lambda|\sigma}\nabla_{AB}\Phi^{|\rho]l}(x)+...\right)\;,
$$
which may be written in the more conventional form
\begin{eqnarray}
{\cal W}^{\alpha\beta}(x,\theta)&=&\Phi^{\alpha\beta}(x)+\theta^A_{\lambda}\left(\Omega^{\alpha\lambda}\Psi^{\beta}_A(x)-\Omega^{\beta\lambda}\Psi^{\alpha}_A(x)-\frac{1}{2}\Omega^{\alpha\beta}\Psi^{\lambda}_A(x)\right)\nonumber\\
&&+\theta^A_{\lambda}\theta^B_{\rho}\left(\left(\Omega^{\alpha\lambda}\Omega^{\beta\rho}-\frac{1}{4}\Omega^{\alpha\beta}\Omega^{\lambda\rho}\right)\,H_{AB}(x)+\Omega^{\alpha\lambda}\nabla_{AB}\Phi^{\beta\rho}(x)-\Omega^{\beta\lambda}\nabla_{AB}\Phi^{\alpha\rho}(x)\right)+{\cal O}(\theta^3)\nonumber
\end{eqnarray}
which is the correct expansion of the superfield. Terms of higher order in $\theta$ can also be found in this way.

\end{document}